\documentstyle[12pt]{article}
\begin{document}
\pagestyle{empty}
\begin{center}

{\large {\bf On the role of a new type of correlated disorder in extended
electronic states in the Thue-Morse lattice}}\\
\vskip 1cm
     Arunava Chakrabarti\footnote{{\bf  Permanent  Address  }:  Scottish
Church College, 1 \& 3 Urquhart Square, Calcutta 700 006, India. }, S. N.
Karmakar and R. K. Moitra\\
{\em Saha Institute of Nuclear Physics, 1/AF, Bidhannagar, Calcutta  700
064, India. }\\
\vskip 1cm
{\bf Abstract }
\end{center}
        A new type of correlated disorder is shown to be responsible
for the appearance of extended electronic states in one-dimensional
aperiodic systems like the Thue-Morse lattice. Our analysis leads to
an understanding of the underlying reason for the extended states in
this system, for which only numerical evidence is available in the
literature so far. The present work also sheds light on the restrictive
conditions under which  the extended states are supported by this
lattice.

\newpage
\pagestyle{plain}

          The  role  of  correlated  disorder  in   producing   extended
electronic  states  in  one-dimensional  disordered  systems  has   been
extensively discussed in the literature by now \cite{ref1}-\cite{ref3}.
The  basic  reason
for the appearance of extended states in such cases  has been traced  to
the existence of a certain type of short range clustering  effect  among
the atoms, first pointed out by Dunlap et al. \cite{ref1}
in their  study  of  a
distribution of random dimers on a host lattice.  It was shown that at a
certain energy value the composite transfer matrix for  a  dimer  offers
identity contribution to the full  transfer  matrix,  so  that  at  this
energy the full disordered lattice effectively  behaves  as  an  ordered
chain. These correlations have later been shown to persist at all length
scales in several quasiperiodic lattices like the copper mean chain  and
the period  doubling  lattice  \cite{ref2},
leading  to  whole  hierarchies  of
extended states.

               Instances of one-dimensional lattices are  however  known
which offer more intriguing  possibilities.  One  such  example  is  the
well-known Thue-Morse aperiodic sequence, in which  there  is  numerical
evidence of extended  states  \cite{ref4}
although  there  is  no  short  range
dimer-type positional correlations in this lattice. As we shall  see  in
this paper, another kind of clustering effect manifests itself  in  this
lattice, which cannot be analysed by the standard method  known  so  far
and developed in  detail  in  Ref.  \cite{ref1} and  \cite{ref2}.
We  may  mention  in  this
connection the recent work of  Lin and Goda \cite{ref5}
concerning a problem  of
hierarchical distribution of potentials on a lattice,  where  they  also
find  an infinite number of extended  electronic  eigenstates,  although
there are  no  dimer-type  correlations.   The  present  work  therefore
provides yet another example of an aperiodic lattice where the  standard
picture of correlated disorder does not work.

In a recent work on the Thue-Morse lattice,  Ryu  et  al.  \cite{ref4}
have provided numerical evidence of the existence of extended electronic
eigenstates at certain specified energies, based on  the  study  of  the
trace map for this lattice. It has been shown that a particular value of
the   trace   remains   unchanged   under   a   renormalisation    group
transformation. By using a detailed  multifractal  analysis  they  found
that the energies associated with  this  trace  value  correspond  to
extended  eigenstates.  This  work,  however,   does  not  provide   any
analytical guideline for discerning  the  extended  character  of  these
eigenfunctions, and the physical mechanism underlying the  behaviour  of
these  eigenfunctions,  independently  of  any  multifractal   analysis,
remains to be clarified. The main purpose of this  work  is  to  provide
this guideline. Additionally, we demonstrate the restrictive  conditions
under which this    lattice  supports  extended  states.  We  show  that
extended states are supported on this lattice  only within the framework
of the  on-site  model;  the  other  commonly  used  model,  namely  the
bond-model,  does  not  support  extended  states,  so  that  our   work
invalidates the findings  of  Ref.  \cite{ref6}  regarding  the  analysis
of  a multi-band model of the Thue-Morse lattice. A rigorous repudiation
of the results of Ryu et al. has recently also been given by Bovier and
Ghez \cite{bov}.

               A Thue-Morse (TM) chain may be built up by starting  with
two letters  $A$ and $B$ and using the inflation rule    $A  \rightarrow
AB$ and $B \rightarrow BA$ repeatedly, where the letters $A$ and $B$ may
be thought of as representing the atoms $A$ and $B$. The  ratio  of  the
numbers of $A$ and $B$ atoms in any generation $\ell$  is
$N_A(\ell)/N_B(\ell) = 1$.
The electronic properties of such a lattice  are
generally modelled by  the  on-site  version  of  the  nearest-neighbour
hamiltonian in the Wannier basis, the Schrodinger equation corresponding
to which is given by
\begin{equation}
(E - \epsilon_n)\psi_n =\psi_{n+1} + \psi_{n-1}  \label{eq1}
\end{equation}
$n$ being the site index, and $\epsilon_n = \epsilon_A$ or $\epsilon_B$,
and the energies are in units of the hopping integral $t$ which has been
set equal to unity.

          The amplitudes of the wave functions at different atomic sites
are determined by using the unimodular transfer matrices of the form
\begin{equation}
M_n = \left (
\begin{array}{cc}
E-\epsilon_n & -1\\
1 & 0
\end{array} \right )~. \label{eq2}
\end{equation}

The amplitude of the wave function at the $n$-th site is then related to
that at the first site by the following matrix product
\[
M_n M_{n-1} \cdots \cdots \cdots M_1~.
\]

          In order to unravel the correlations that are responsible  for
the extended states in a Thue-Morse chain,  we  first  recapitulate  the
structural peculiarities of this lattice. At the very  basic  level,  we
may regard the atoms $A$ and $B$ to be the basic building blocks of  the
TM lattice; at the next level the pairs $AB$ and $BA$ may be regarded as
the basic starting elements on which the  repeated  application  of  the
Thue-Morse  inflation  rules  generates  the  whole  lattice.   We   may
successively consider the pair of quadruplets  $ABBA$  and  $BAAB$,  the
pair  $ABBABAAB$  and  $BAABABBA$,  the  pair   $ABBABAABBAABABBA$   and
$BAABABBAABBABAAB$  etc. as  the  starting  blocks  for  generating  the
Thue-Morse chain by applying the appropriate inflation rules. Evidently,
there are no dimer or higher order atomic clustering in this lattice. It
thus becomes necessary to analyse  the  products  of  transfer  matrices
corresponding to these strings of atoms  in  order  to   understand  the
conditions under which extended states are supported by the TM lattice.

                   To handle such long products of matrices we resort to
the following device. Let us  begin  by  resolving   the  $2  \times  2$
matrices $M_n$ in a basis formed by the $2  \times  2$  identity  matrix
$I$ and the three Pauli matrices  $\sigma_x$, $\sigma_y$ and $\sigma_z$.
We then have the  expressions for $M_A$ and $M_B$ in the following form:
\begin{equation}
\begin{array}{lll}
M_A & = & \alpha_A I + \beta_A \sigma_x + \gamma_A \sigma_y  +  \delta_A
\sigma_z\\
\\ \nonumber
M_B & = & \alpha_B I + \beta_B \sigma_x + \gamma_B \sigma_y  +  \delta_B
\sigma_z
\end{array} \label{eq3}
\end{equation}
where   $\alpha_{A(B)}    =    \delta_{A(B)}    =(E-\epsilon_{A(B)})/2$,
$\beta_{A(B)}=0$ and $\gamma_{A(B)}= -i$.

       Using  Eqs.(\ref{eq3})  and  the  commutation  rules  for
the   matrices
$\sigma_x$, $\sigma_y$ and $\sigma_z$, we may easily resolve the  matrix
products $M_A M_B$ and $M_B M_A$ to obtain
\begin{equation}
\begin{array}{lll}
M_A M_B & = & \alpha_1 I  +  \beta_1  \sigma_x  +  \gamma_1  \sigma_y  +
\delta_1 \sigma_z\\
\\       \nonumber
M_B M_A & = & \alpha_1 I  -  \beta_1  \sigma_x  +  \gamma_1  \sigma_y  +
\delta_1 \sigma_z
\end{array} \label{eq4}
\end{equation}
where  $\alpha_1  =  2  \alpha_A  \alpha_B  -1$,  $\beta_1  =   \alpha_B
-\alpha_A$, $\gamma_1 = -i(\alpha_A  +  \alpha_B)$  and  $\delta_1  =  2
\alpha_A \alpha_B$. We  notice that the  coefficients  of  the  matrices
$I$, $\sigma_y$ and $\sigma_z$ are the same for both the  products  $M_A
M_B$ and $M_B M_A$ in Eqs.(\ref{eq4}),
while the  coefficient  of  $\sigma_x$  in
these expressions appear with  opposite  signs.  Thus  if  we  set  this
coefficient, ie. $\beta_1 =0$, then the products $M_A M_B$ and $M_B M_A$
become identical, and the  system  effectively  reduces  to  an  ordered
binary chain of alternating $A$ and $B$ atoms.  However,  it  turns  out
that the condition $\beta_1 = 0$ can only be satisfied with  $\epsilon_A
= \epsilon_B$.  We thus have to proceed to the next stages, and  compare
the pairs of matrix products $M_A M_B M_B M_A$ and $M_B  M_A  M_A  M_B$,
the products $M_A M_B M_B M_A M_B M_A M_A M_B$ and $M_B M_A M_A M_B  M_A
M_B M_B M_A$ and so on.

               Using the  basic  results  for  the  resolutions  of  the
matrices $M_A$ and $M_B$ (Eqs.(\ref{eq3})) and
for the  products  $M_A  M_B$  and
$M_B M_A$ (Eqs.(\ref{eq4}))  repeatedly,
we can  easily  find  the  forms  of  the
longer matrix products mentioned above. Interestingly, the
pair of products of matrices $M_A M_B M_B M_A M_B M_A  M_A  M_B$  \ldots
and $M_B M_A M_A M_B M_A M_B M_B M_A$ \ldots each with $2^n$ elements show
a surprising regularity having either of the following forms
\begin{equation}
\begin{array}{lll}
M_A M_B M_B M_A M_B M_A M_A M_B \cdots &  =  &  \alpha_n  I  +  \gamma_n
\sigma_y + \delta_n \sigma_z \\
\\
M_B M_A M_A M_B M_A M_B M_B M_A \cdots & =  &  \alpha_n  I  +  \gamma'_n
\sigma_y + \delta'_n \sigma_z
\end{array}  ~~\mbox{\rm ($n$ even)} \label{eq5}
\end{equation}
and
\begin{equation}
\begin{array}{lll}
M_A M_B M_B M_A M_B M_A M_A M_B \cdots  &  =  &  \alpha_n  I  +  \beta_n
\sigma_x + \gamma_n \sigma_y + \delta_n \sigma_z \\
\\
M_B M_A M_A M_B M_A M_B M_B M_A \cdots  &  =  &  \alpha_n  I  -  \beta_n
\sigma_x + \gamma_n \sigma_y + \delta_n \sigma_z
\end{array} ~~\mbox{\rm ($n$ odd)}. \label{eq6}
\end{equation}

          It follows that  the  matrix  products
$M_A M_B M_B M_A M_B M_A M_A M_B \cdots$ and $M_B M_A M_A  M_B  M_A  M_B
M_B M_A \cdots$ can never be made equal at  some  value  of  the  energy
for even values of  $n$
because their expansions given in Eqs.(\ref{eq5}) differ
in  two  coefficients
$\gamma$ and $\delta$. On the other hand, for odd values of $n$,
Eqs.(\ref{eq6})
show that the matrix products become equal to each other if  $\beta_n  =
0$. Thus the energy values for which $\beta_n = 0$  ($n$  odd)  are  the
ones for which the composite transfer matrices for  of  the  strings  of
atoms  $ABBABAAB  \cdots$  and   $BAABABBA   \cdots$   offer   identical
contributions to the full transfer matrix for the whole chain. Since the
Thue-Morse chain may be regarded as built out of these composite blocks,
the full transfer matrix for the whole chain now consists of  a  product
of identical unimodular $2 \times 2$ matrices $M$, each corresponding to
a block of atoms $ABBABAAB \cdots$ or $BAABABBA  \cdots$.  This  however
does not mean that one  necessarily  has  extended  states  at  all  the
energies obtained by setting $\beta_n = 0$.  Several of  these  energies
in fact lead to diverging, and therefore disallowed  functions,  and  a
careful analysis has to be made to determine  precise  conditions  under
which the string of such identical matrices leads to extended states, as
the length of the chain becomes arbitrarily large.

          We  first  give  the  recursion   relations   connecting   the
coefficients $\alpha$, $\beta$, $\gamma$ and $\delta$ for two successive
odd values of $n$ ($n \geq  3$)  which  are  easily  obtained  by  using
Eqs.(\ref{eq5}) and Eqs.(\ref{eq6}). These are
\begin{equation}
\begin{array}{lll}
\alpha_n  &  =  &  \alpha_{n-2}^4  +  \beta_{n-2}^4   +   \gamma_{n-2}^4
+\delta_{n-2}^4 + 2 (\gamma_{n-2}^2 + \delta_{n-2}^2) (3  \alpha_{n-2}^2
+ \beta_{n-2}^2)\\
& & + 2 (\gamma_{n-2}^2  \delta_{n-2}^2   -  \alpha_{n-2}^2
\beta_{n-2}^2)\\
\\
\beta_n  &  =  &  8  \alpha_{n-2}   \beta_{n-2}   (   \gamma_{n-2}^2   +
\delta_{n-2}^2 )\\
\\
\gamma_n & = & 4 ( \alpha_{n-2}^2 -  \beta_{n-2}^2  +  \gamma_{n-2}^2  +
\delta_{n-2}^2 ) \alpha_{n-2} \gamma_{n-2}\\
\\
\delta_n & = & 4 ( \alpha_{n-2}^2 -  \beta_{n-2}^2  +  \gamma_{n-2}^2  +
\delta_{n-2}^2 ) \alpha_{n-2} \delta_{n-2}
\end{array} \label{eq7}
\end{equation}

          From Eqs.(\ref{eq7}) we find that the condition
$\beta_n =  0$  implies
either of the following conditions:
\begin{eqnarray}
\alpha_{n-2} & = & 0\\ \label{eq8}
\beta_{n-2} & = & 0\\   \label{eq9}
\gamma_{n-2}^2 + \delta_{n-2}^2 & = & 0. \label{eq10}
\end{eqnarray}

     First, it can be easily  shown  with  the  help  of  the  recursion
relations (\ref{eq7}) that the value of $\alpha_n = 1$
in each  of  these  cases.
Among the three conditions Eqs.(8-10),
only the condition  $\alpha_{n-2}
= 0$ automatically leads to $\gamma_n=0$ and $\delta_n = 0$,  while  the
two other conditions (9) and (10) generally
yield  non-zero  values  of
$\gamma_n$ and  $\delta_n$.  As  will  be  shown  presently,  the  first
condition always leads  to  allowed  values  of  the  energies  for  the
extended states in the TM lattice. The energies obtained  from  the  two
other conditions,  Eqs.(9)  and  (10),
generally  lead  to  asympotically
diverging functions, except in special  circumstances  to  be  discussed
later.

     To carry out this analysis, let us express the matrix  $M$  at  any
level in the form ( suppressing the index $n$ )
\[
M = \left (
\begin{array}{cc}
\alpha + \delta & \beta - i\gamma \\
\beta + i\gamma & \alpha -\delta
\end{array} \right )~.
\]

           $\alpha$ can be identified with the quantity  $(1/2) \mbox{\rm
Tr}~M$.  We now recall the well-known result \cite{ref7} for the powers of
$2 \times 2$ unimodular matrices $M$
\begin{equation}
M^m  =  U_{m-1}(x) M - U_{m-2}(x) I \label{eq11}
\end{equation}
for $m \geq 2$. Here $U_m(x)$ is the Chebyshev polynomial of the second
kind,  and  $x$
is the quantity $(1/2) \mbox{\rm  Tr}~M = \alpha$. The function $U_m(x)$
is defined as $\sin (m+1)\theta /\sin \theta$ with $\cos \theta =x$. In the
present discussion $\alpha_n  = 1$ in every case, for which $U_m(1)$ has
the limiting value $(m + 1)$. We then have the  following  expression  for
$M^m(x=1)$:
\begin{equation}
M^m = \left (
\begin{array}{cc}
m(\alpha +\delta) -m+1 & m(\beta - i\gamma) \\
m(\beta + i\gamma) & m(\alpha -\delta) -m +1
\end{array} \right )~. \label{eq12}
\end{equation}

          Let us  now  choose,  without  any  loss  of  generality,  the
amplitudes of the wavefunctions at the zero-th and the first site to  be
$0$ and $1$ respectively. With the aid of Eq.(\ref{eq1})
it then becomes  simple
to express the amplitudes at the $(m + 1)$-th and the $m$-th sites as,
\[
\psi_{m+1}  = m ( \alpha + \delta - 1) + 1
{}~~~~\mbox{\rm and}~~~~
\psi_m = m ( \beta +i \gamma )~.
\]

          We  now  observe  that  the  finiteness  of  $\psi_m$  at   $m
\rightarrow \infty$ is assured only when $\alpha + \delta =  1$  and
$\beta + i \gamma = 0$ simultaneously. Now, as we have already seen,
having  $\alpha_{n-2}  =  0$  at  some  stage  automatically  makes  both
$\gamma_n$ and $\delta_n$ equal to zero, in addition to making $\alpha_n
= 1$. Therefore, the criterion for the finiteness of the wavefunction  at
$m \rightarrow \infty$ is satisfied in this case. On  the  other  hand,
since neither of the conditions  Eqs.(9) and (10)
yield  $\gamma_n$  and
$\delta_n$ equal to zero, so  that  even  though  $\alpha_n  =  1$,  the
amplitudes diverge  with  increasing  $m$. The  energies
obtained from these conditions may either not lead to proper eigenstates
for the system or they may correspond to eigenstates with power law
growth at infinity.
An exception occurs when in condition (10)
the  quantities
$\gamma_{n-2}$ and $\delta_{n-2}$ possess common factors, so that energy
values exist for which $\gamma_{n-2}$  and  $\delta_{n-2}$  become  zero
simultaneously. In this case  $\gamma_n$ and  $\delta_n$  become
zero and we again have extended states at these energies.

          Hence we arrive at the  condition  $\alpha_{n-2}  =  0$
that gives rise to a  set  of  energy  eigenvalues  which  unambiguously
correspond to extended eigenstates for the  infinite  TM  lattice.  Each
block $M_A M_B M_B M_A M_B M_A M_A M_B \cdots$ or  $M_B  M_A  M_A
M_B M_A M_B M_B M_A \cdots$ of size $2^n$ ($n$ odd ) at  these  energy
eigenvalues
becomes an identity matrix and therefore the total  transmission  across
the chain becomes unity at these energies. For all other  energy  values
arising out of the conditions (9)  and  (10),
which  of  course  also
correspond to the basic condition $\beta_n =  0$,  $\psi_m$  grows  with
$m$, and therefore they are not permissible.  As  mentioned  above,  the
only exception occurs when $\gamma_{n-2}$ and $\delta_{n-2}$ have common
factors and therefore vanish for the same value of  the  energy  at  the
same time.

               Although it is straighforward  to  obtain  the  algebraic
expressions  for  the  energy  values  obtained   from   the   condition
$\alpha_{n-2} = 0$, we calculate  instead the energies  for  a  specific
choice of the parameter values, namely $\epsilon_A = -\epsilon_B = 0.5$.
We then obtain the following energy values for the first two levels :
\[
\begin{array}{lllll}
n & = & 3 & : & \pm 1.5 \\
n & = & 5 & : & \pm 2.094225204000, \pm 1.706503280648,\\
  &   &   &   & \pm 1.260097834749, \pm 0.337965671230~.
\end{array}
\]

          For the sake of illustration we display in Fig.1 the variation
of $|\psi|^2$ with the site-index $n$ for a  few  selected  energies.  The
signature of the  underlying  Thue-Morse  aperiodicity  is  particularly
apparent  in  Figs.1(a), 1(c) and 1(d);  all  the   wave-functions   are
lattice-like. The general characteristic of these functions is that with
increasing $n$, amplitudes  tend  to  cluster  around  groups  of  sites
separated by islands where the amplitudes have very low values,  a  fact
which was also pointed out by Ryu et al. \cite{ref6}.

               The spirit in which the present analysis has been carried
out corresponds to the real space renormalisation group point  of  view.
At any length scale corresponding to an odd value of $n$, Eqs.(\ref{eq7})
define
effective $A$ and $B$ atoms arranged in  a  TM  sequence.  The  extended
character of the eigenfunctions discussed in this paper owe their origin
to a peculiar short range correlation among the atoms  at  every  length
scale, in contrast to  the  usual  dimer-type  short  range  correlation
discussed so far in the literature.

                    The analysis so far has been made on  the  basis  of
the on-site model of the TM lattice. In dealing with the  pure  transfer
model in which we have two different  hopping  strengths  for  the  long
($L$) and short ($S$) bonds, it turns out that the full transfer  matrix
for the chain is composed  of  a  product  of  four  different  transfer
matrices corresponding to the $LL$, $LS$, $SL$ and $SS$ pairs.  This  is
in contrast to the unique pair of $2 \times 2$ matrices that can act  as
``building blocks" for the entire TM chain in the on-site case. The  four
basic transfer matrices do not  follow  the  TM  sequence  and  form  an
intractable complex pattern, so that the  recursive  application  of
the TM inflation rule is not  meaningful  in  this  case.  It  therefore
follows that the analysis of  strings  of  transfer  matrices  based  on
Eqs.(\ref{eq5}) and (\ref{eq6}) is no longer possible,
and thus  there  are  no  extended
states in the transfer model. We may point out that the matrix recursion
relation employed by Ryu et al. \cite{ref6},
$M_l = M_{l-1}  \overline{M}_{l-1}$
is incorrect for the transfer model, because  the string of matrices  do
not take into account the atomic site flanked by a pair  of  $S-S$  bond
and a pair of $L-L$ bond on both sides.  Here,  $M_l$  is  the  transfer
matrix for the $l$-th generation chain, and $\overline{M}_l$ is the same
matrix $M_l$ with the roles of $t_L$ and $t_S$ interchanged. It  can  be
verified easily that the above mapping does not generate the correct  TM
chain in the transfer  model.  Thus  the  analysis  of  Ref.\cite{ref6}
of  the
multiband model based on the above mapping  is not valid. In  fact,  for
mixed models involving variations in both site-energies and the hopping,
we still have four different transfer matrices, and again there  are  no
extended states.

               To conclude, we  have  demonstrated  a  new  mechanism
through which short  range  correlations  lead  to  extended  states  in
aperiodic one dimensional chains such as the TM lattice.  For  this,  we
provide a way of analysis which hopefully will be found useful
in problems of similar kind.

\newpage
\noindent
{\bf Figure Caption}
\begin{itemize}
\item[Fig.1]
Plot of $|\psi_n|^2$ versus site number $n$ for a TM sequence with $256$
sites. Here $\epsilon_A = -\epsilon_B = 0.5$ with energies measured in
units of hopping matrix element $t$. Figs.(a), (b), (c) and (d) correspond
respectively to energies $1.5$, $0.337965671230$, $-1.260097834749$ and
$-1.995507972823$.
\end{itemize}

\newpage
\begin{thebibliography}{20}
\bibitem{ref1} D. H. Dunlap, H. L. Wu  and P. Phillips, Phys. Rev. Lett.
{\bf 65}, 88 (1990);
H. L. Wu and P. Phillips, Phys. Rev. Lett. {\bf 66}, 1366 (1991)
\bibitem{ref2} S. Sil, S. N. Karmakar, R. K. Moitra and A. Chakrabarti,
Phys. Rev. B {\bf 48}, 4192 (1993);
A. Chakrabarti, S. N. Karmakar and R. K. Moitra, Phys. Rev. B {\bf 50},
13276 (1994)
\bibitem{ref3} A. Sanchez, E. Macia and E. Dominguez-Adame, Phys. Rev. B
{\bf 49}, 147 (1994)
\bibitem{ref4} C. S. Ryu, G. Y. Oh and M. H. Lee, Phys. Rev. B {\bf  46},
5162 (1992)
\bibitem{ref5} Z. Lin and M. Goda, J. Phys. A: Math. Gen.
{\bf 26}, L217 (1993)
\bibitem{ref6} C. S. Ryu, G. Y. Oh and M. H. Lee, Phys. Rev. B {\bf  48},
132 (1993)
\bibitem{bov} A. Bovier and J. M. Ghez (private communication)
\bibitem{ref7} A. Cayley, Philos. Trans. Roy. Soc. (London) {\bf 148},
17 (1858); F. Abeles, Ann. de Physique {\bf 5}, 777 (1950); for a recent
presntation see J. M. Luck, ``Aperiodic structures: geometry,
diffraction spectra and physical properties" in {\em Fundamental
Problems in Statistical Mechanics VIII}, Elsevier (1994)
\end{thebibliography}
\end{document}